\pdfoutput=1
\documentclass[sigconf,natbib=true]{acmart}

\usepackage{multirow}
\usepackage{subcaption}
\usepackage[inline]{enumitem}
\usepackage{xspace}
\usepackage{fontawesome5}

\definecolor{headlineourmain}{HTML}{4B2E83}
\definecolor{headlineourlight}{HTML}{EFE8F5}
\definecolor{headlineposgreen}{HTML}{2E7D32}
\definecolor{headlinenegred}{HTML}{C62828}
\definecolor{headlinebbbluelt}{HTML}{F1F6FC}
\definecolor{headlinesigorangelt}{HTML}{FFE8D0}

\newcommand{\swatch}[1]{%
  {\fboxsep=0pt\fboxrule=0.3pt%
   \fcolorbox{black!40}{#1}{\rule{0pt}{1.3ex}\hspace{1.3ex}}}%
}

\definecolor{takeawaybg}{HTML}{EFE8F5}
\definecolor{takeawaybord}{HTML}{4B2E83}
\newcommand{\takeaway}[1]{%
  \par\noindent
  \fcolorbox{takeawaybord}{takeawaybg}{%
    \begin{minipage}{\dimexpr\columnwidth-2\fboxsep-2\fboxrule\relax}%
      \small\textbf{\color{takeawaybord}Takeaway.}\ #1%
    \end{minipage}%
  }%
  \par%
}

\AtBeginDocument{%
  }

\AtBeginDocument{%
  \hypersetup{%
    colorlinks=true,
    citecolor=headlineourmain,
    linkcolor=headlineourmain,
    urlcolor=headlineourmain,
  }%
}

\newcommand{\system}{Dual\-Spec\-tral\-CF\xspace}
\newcommand{\sysCheby}{Dual\-Spec\-tral\-CF-Cheby\xspace}
\newcommand{\sysGF}{Dual\-Spec\-tral\-CF-GF\xspace}
\newcommand{\sysTurbo}{Dual\-Spec\-tral\-CF-Tur\-bo\xspace}
\newcommand{\RR}{\mathbb{R}}
\newcommand{\II}{\mathcal{I}}
\newcommand{\UU}{\mathcal{U}}
\newcommand{\Epos}{\mathcal{E}^+}
\newcommand{\Eneg}{\mathcal{E}^-}
\newcommand{\Rtil}{\tilde{\mathbf{R}}}
\newcommand{\Lstar}{\mathbf{L}^*}
\newcommand{\Lsign}{\mathbf{L}^{\pm}}
\newcommand{\Ditem}{\mathbf{D}_\II}
\newcommand{\Duser}{\mathbf{D}_\UU}

\setcopyright{none}
\renewcommand\footnotetextcopyrightpermission[1]{}
\copyrightyear{2026}
\acmYear{2026}
\acmConference[CIKM '26]{The 35th ACM International Conference on Information and Knowledge Management}{November 7--11, 2026}{Rome, Italy}

\begin{document}

\title[DualSpectralCF]{DualSpectralCF: Training-Free Sign-Aware Spectral Collaborative Filtering}
\titlenote{Accepted at the 35th ACM International Conference on Information and Knowledge Management (CIKM '26), November 7--11, 2026, Rome, Italy. This is the authors' preprint version.}

\author{Guanqun Yang}
\email{guanqun.yang@outlook.com}
\affiliation{%
  \institution{Stevens Institute of Technology}
  \city{Hoboken}
  \state{NJ}
  \country{USA}
}

\author{Tong Qi}
\email{tongqi0110@gmail.com}
\affiliation{%
  \institution{University of Maryland, College Park}
  \city{College Park}
  \state{MD}
  \country{USA}
}

\author{Xiaoxue Han}
\email{xhan26@stevens.edu}
\affiliation{%
  \institution{Stevens Institute of Technology}
  \city{Hoboken}
  \state{NJ}
  \country{USA}
}

\renewcommand{\shortauthors}{Yang et al.}

\begin{abstract}
Real-world recommendation platforms routinely collect explicit negative feedback such as 1-star reviews, hate-button clicks, distrust between users, and very-low watch-ratio videos.
Learned sign-aware recommenders exploit this signal for clear accuracy gains, but only at the cost of gradient-based training.
In parallel, a line of training-free spectral collaborative filtering methods matches or beats learned graph recommenders at a fraction of the cost, yet operates on positive interactions alone.
We bridge these two lines with \system, a training-free framework of two components that attach to any spectral backbone of the form $\hat{\mathbf{r}}_u = F(\mathbf{M})\,\mathbf{r}_u$: a signed input signal $\mathbf{r}_u^{\pm}$ that encodes the user's explicit dislikes, and a signed item-item operator $\mathbf{M}^{\pm}$ that blends like-together and dislike-together similarity.
The framework is backbone-agnostic and adds just two scalar hyperparameters.
We instantiate \system on ChebyCF, GF-CF, and Turbo-CF, and evaluate on five sign-aware benchmarks: every instance matches or beats its unsigned backbone on all 5 datasets, with Recall@20 lifts up to +32.6\% with backbone-specific $(\gamma, \kappa)$ tuning and +1.9\% to +16.0\% for \sysCheby at the fixed default $(\gamma{=}{-}0.5, \kappa{=}0.1)$, and the family runs 7.7 to 155.3$\times$ faster than SIGformer while reaching 70.7\% to 90.7\% of its accuracy.
Sign-awareness helps most for cold-start users, with up to +29.2\% Recall@20 on Epinions users with 1 to 5 training items.

\par\noindent\textbf{\faGithub\ Code:} \url{https://github.com/guanqun-yang/DualSpectralCF}
\end{abstract}



\maketitle

\section{Introduction}
\label{sec:intro}

Collaborative filtering (CF) is a foundational technique in modern recommender systems: it predicts a user's preference for unseen items by aggregating the past behavior of other users with similar interaction patterns.
Over the past decade, CF has moved from classical matrix-factorization-style systems to graph-based formulations that view the user-item interaction matrix as a bipartite graph~\citep{Wang2019NeuralGraphCollaborative, He2020LightGCNSimplifyingPowering, Cai2023LightGCLSimpleEffective, Wu2022GraphNeuralNetworks, Wang2020GraphLearningApproaches, Zhang2023RecommendingGraphsComprehensive}.
Unlike matrix factorization, which captures only single-step user-item associations, the bipartite graph lets evidence travel along multi-hop paths such as $u_1 \!\to\! i_1 \!\to\! u_2 \!\to\! i_2$, recommending $i_2$ to $u_1$ through a user $u_2$ who shares an item with $u_1$ even when the two share no other item~\citep{Wang2019NeuralGraphCollaborative, He2020LightGCNSimplifyingPowering}.

The standard graph-based CF pipeline, however, considers \emph{positive} interactions only: a user-item edge exists when the user interacted with the item, and is otherwise absent.
Many production platforms also collect \emph{explicit negative} feedback, such as 1-star Amazon reviews~\citep{Ni2019JustifyingRecommendationsUsing}, hate-button clicks in KuaiRand~\citep{Gao2022KuaiRandUnbiasedSequential}, user-to-user distrust in Epinions~\citep{Tang2012ETrustUnderstandingTrust}, and very-low watch-ratio videos in KuaiRec~\citep{Gao2022KuaiRecFullyobservedDataset}.
Unlike an absent edge in the user-item interaction graph, which may simply mean the user has not yet seen the item, this signal reveals what the user actively dislikes.
A line of \textbf{sign-aware} recommenders has emerged to use both signs: SiReN~\citep{Seo2022SiReNSignAwareRecommendation} and PANE-GNN~\citep{Liu2023PANEGNNUnifyingPositive} split the bipartite graph by sign and fuse the halves through learned attention, NFARec~\citep{Wang2024NFARecNegativeFeedbackAware} models signed feedback sequences with a Transformer Hawkes process, and SIGformer~\citep{Chen2024SIGformerSignawareGraph} and DFGNN~\citep{Wu2024DFGNNDualfrequencyGraph} fold signed spectral information into a Transformer and a dual-frequency GNN, respectively.
More recent signed models unify the dual encoders into a single encoder (LSGRec~\citep{Liu2026UnifiedModelingPositive}) or apply a Gegenbauer-basis spectral convolution to signed bipartite graphs (GegenNet~\citep{Wang2025GegenNetSpectralConvolutional}).
These methods deliver clear accuracy gains from the additional signal, but each is gradient-based and considerably more expensive than the closed-form spectral methods that our training-free framework, \system, builds on.

A complementary line of work treats each user's interaction row as a signal on the item-item similarity graph and produces recommendations by passing it through a low-pass filter that suppresses noisy high-frequency components while retaining smooth low-frequency ones.
The pipeline is closed-form and \emph{training-free}, yet it matches or beats learned graph neural network (GNN) recommenders at a fraction of the cost.
GF-CF~\citep{Shen2021HowPowerfulGraph} showed that this signal processing view recovers the accuracy of GCN-based CF; later work made it faster or more flexible by replacing the eigendecomposition with polynomial filters (Turbo-CF~\citep{Park2024TurboCFMatrixDecompositionFree}, PolyCF~\citep{Qin2024PolyCFOptimalSpectral}, and ChebyCF~\citep{Kim2025GraphSpectralFiltering}) or by adding a sharpening step that emphasizes top items (BSPM~\citep{Choi2023BlurringSharpeningProcessModels}).
Closed-form positive-only methods such as EASE~\citep{Steck2019EmbarrassinglyShallowAutoencoders} and PSGE~\citep{DAmico2023PureSpectralGraph} likewise avoid gradient training, while SpectralCF~\citep{Zheng2018SpectralCollaborativeFiltering}, JGCF~\citep{Guo2023ManipulatingSignalsUserItem}, and GDE~\citep{Peng2022LessMoreReweighting} vary the polynomial basis or reweight spectral features.
However, this spectral family has so far focused on positive interactions and has not been extended to incorporate explicit negative feedback. \system closes this gap.

We build \system from two components.
A standard spectral CF method scores items as $\hat{\mathbf{r}}_u = F(\mathbf{M})\,\mathbf{r}_u$, where $\mathbf{r}_u$ is the user's interaction row and $F(\mathbf{M})$ is a fixed filter applied to an item-item operator $\mathbf{M}$.
Our first component replaces the binary input $\mathbf{r}_u$ with a \emph{signed input signal} $\mathbf{r}_u^{\pm}$ that also encodes the user's explicit dislikes, controlled by a scalar $\gamma$.
Our second component replaces $\mathbf{M}$ with a \emph{signed item-item operator} $\mathbf{M}^{\pm}$ that combines two kinds of item-to-item similarity, items that users commonly like together and items that users commonly dislike together, controlled by a scalar $\kappa$.
Each component thus changes a different part of the same scoring equation $\hat{\mathbf{r}}_u = F(\mathbf{M})\,\mathbf{r}_u$ (the input $\mathbf{r}_u$ for one, the operator $\mathbf{M}$ for the other), so the two are independent and complementary.
Together they attach to any backbone of this form, keeping the framework backbone-agnostic and training-free while adding exactly two hyperparameters beyond those the backbone already uses.
We instantiate \system on three heterogeneous backbones to obtain \sysCheby (on ChebyCF~\citep{Kim2025GraphSpectralFiltering}), \sysGF (on GF-CF~\citep{Shen2021HowPowerfulGraph}), and \sysTurbo (on Turbo-CF~\citep{Park2024TurboCFMatrixDecompositionFree}).
The default $(\gamma{=}{-}0.5,\ \kappa{=}0.1)$ works for \sysCheby on 4 of 5 datasets; \sysGF and \sysTurbo benefit from per-dataset $\kappa$.

\begin{figure*}[t]
  \centering
  \includegraphics[width=\textwidth]{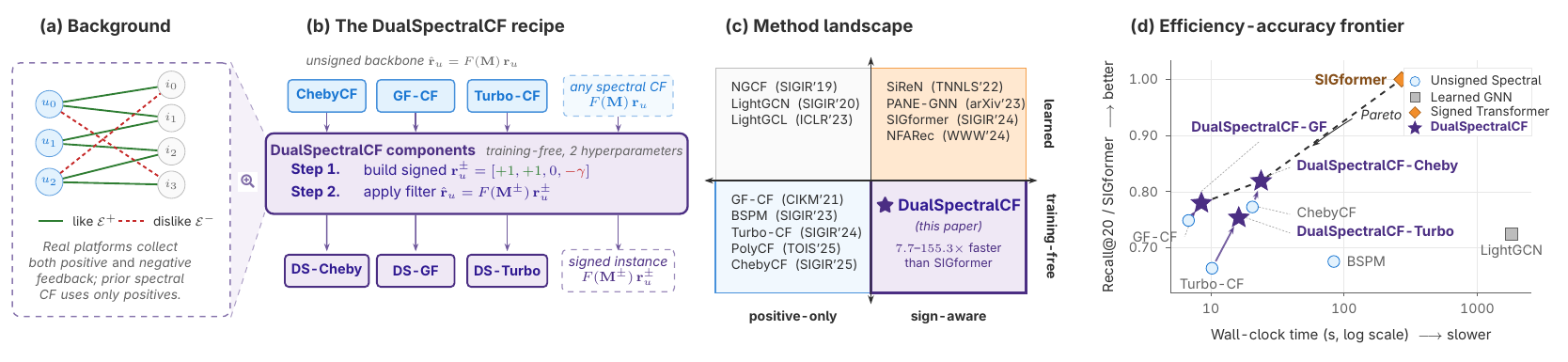}
  \caption{
  (a)~Explicit feedback has positive ({\color{headlineposgreen}\rule[0.6ex]{1.5ex}{0.4pt}}) and negative ({\color{headlinenegred}\rule[0.6ex]{1.5ex}{0.4pt}\hspace{-1.3ex}\rule[0.6ex]{1.5ex}{0.4pt}}) edges; \system assembles the signed row $\mathbf{r}_u^{\pm}$ (positives $+1$, negatives $-\gamma$) and applies the filter $F(\mathbf{M}^{\pm})$ to score items $\hat{\mathbf{r}}_u$.
  (b)~The two components ($\mathbf{r}_u^{\pm}$, $\mathbf{M}^{\pm}$) attach to any backbone $\hat{\mathbf{r}}_u = F(\mathbf{M})\,\mathbf{r}_u$, adding only $\gamma, \kappa$; we instantiate on ChebyCF, GF-CF, and Turbo-CF.
  (c)~\system fills the previously empty sign-aware training-free cell (training-free \swatch{headlinebbbluelt}, SIGformer \swatch{headlinesigorangelt}, ours \swatch{headlineourlight}).
  (d)~On the efficiency-accuracy frontier (horizontal: wall-clock time, log scale; vertical: Recall@20 normalized to SIGformer), each \system instance (stars) improves on its unsigned backbone, \sysGF and \sysCheby lie on the training-free Pareto frontier, and the family runs 7.7 to 155.3$\times$ faster than SIGformer.}
  \label{fig:pipeline}
\end{figure*}

We evaluate \system against four research questions in \S\ref{sec:experiment}, and our contributions are:
\begin{itemize}[leftmargin=*, topsep=2pt, partopsep=0pt, itemsep=3pt]
  \item \textbf{Accuracy.} \system matches or beats every unsigned spectral backbone on all 5 datasets and beats LightGCN on 4 of 5, reaching 70.7\% to 90.7\% of SIGformer's Recall@20 (Table~\ref{tab:main-brief}).
  \item \textbf{Generality.} The two components transfer across different spectral backbones; the default $(\gamma, \kappa){=}({-}0.5, 0.1)$ works without modification for \sysCheby on 4 of 5 datasets, and \sysGF/\sysTurbo need only per-dataset $\kappa$ (Table~\ref{tab:transfer-brief}).
  \item \textbf{Efficiency.} \sysGF and \sysCheby sit on the training-free Pareto frontier, and the family runs 7.7 to 155.3$\times$ faster than SIGformer at 16.1\% to 58.8\% time overhead over the unsigned backbones (Figure~\ref{fig:pipeline}(d), Table~\ref{tab:efficiency-brief}).
  \item \textbf{Cold-start.} Sign-awareness most helps cold-start users, whose sparse positive histories make the negative channel proportionally more informative (\S\ref{sec:rq4}).
\end{itemize}

\section{\system}
\label{sec:method}

As introduced in \S\ref{sec:intro}, \system attaches two components to the prediction equation $\hat{\mathbf{r}}_u = F(\mathbf{M})\,\mathbf{r}_u$: a \emph{signed input signal} $\mathbf{r}_u^{\pm}$ that replaces $\mathbf{r}_u$, and a \emph{signed item-item operator} $\mathbf{M}^{\pm}$ that replaces $\mathbf{M}$.
We first describe the common abstraction that covers the spectral CF backbones we build on (\S\ref{sec:method-backbone}), then formalize the two components at the framework level (\S\ref{sec:method-signal}, \S\ref{sec:method-laplacian}).

\subsection{A Common Spectral Backbone Abstraction}
\label{sec:method-backbone}

Let $\mathbf{R} \in \{0,1\}^{|\UU| \times |\II|}$ denote the binary user-item interaction matrix and let $\mathbf{r}_u \in \{0,1\}^{|\II|}$ be user $u$'s interaction row.
Define the symmetrically normalized interaction matrix $\Rtil = \Duser^{-1/2} \mathbf{R}\, \Ditem^{-1/2}$, the item-item Laplacian $\Lstar = \mathbf{I} - \Rtil^{\top}\Rtil$, and the item-item similarity $\hat{\mathbf{P}} = \Rtil^{\top}\Rtil$.
Every training-free spectral CF backbone we consider can be written as $\hat{\mathbf{r}}_u = F(\mathbf{M};\,\text{hp})\,\mathbf{r}_u$, where $\mathbf{M}$ is an item-item operator (either $\Lstar$ or $\hat{\mathbf{P}}$), $F$ is a scalar-valued graph filter applied to $\mathbf{M}$, and $\text{hp}$ denotes backbone-specific hyperparameters.
The three backbones we evaluate differ only in the form of $F$:
{\small\begin{align}
  F_{\text{GF}}(\Lstar) &= \hat{\mathbf{P}} + \alpha\, H_I(\Lstar;\eta), \label{eq:gfcf}\\
  F_{\text{Turbo}}(\hat{\mathbf{P}}) &= \sum_{k=1}^{K} a_k\, \hat{\mathbf{P}}^{\circ s\,k}, \label{eq:turbocf}\\
  F_{\text{Cheby}}(\Lstar) &= \Ditem^{\beta}\bigl(H_C(\Lstar;\phi) + \alpha H_I(\Lstar;\eta)\bigr)\Ditem^{-\beta}. \label{eq:chebycf}
\end{align}}
where $\circ s$ denotes a Hadamard power, $H_C$ and $H_I$ are Chebyshev and ideal-pass filters with $H_C(\tilde{\mathbf{L}}) = \sum_{k=0}^{K} c_k T_k(\tilde{\mathbf{L}})$, and the coefficients $c_k$ come from Chebyshev interpolation of a plateau transfer function at the Chebyshev nodes.

\subsection{Component A: Signed Input Signal}
\label{sec:method-signal}

We replace the binary row $\mathbf{r}_u$ with a signed signal that includes negative feedback:
\begin{equation}
  \mathbf{r}_u^{\pm}[i] = \begin{cases}
    +1       & \text{if } (u,i) \in \Epos, \\
    -\gamma  & \text{if } (u,i) \in \Eneg, \\
    0        & \text{otherwise,}
  \end{cases}
  \label{eq:signed-signal}
\end{equation}
where $\gamma \in \RR$ is a single new hyperparameter and $\Epos$, $\Eneg$ denote the positive and negative interaction edges.
The backbone's filter $F(\mathbf{M})$ is unchanged; only the input signal is replaced, so component A applies to \emph{any} backbone of the prediction template above.

\paragraph{Why $\gamma < 0$ Works.}
The natural expectation is that $\gamma > 0$ should be optimal, treating negative feedback as evidence \emph{against} spectrally similar items: a disliked item at position $i$ would carry weight $-\gamma < 0$, which the low-pass filter propagates to nearby items, lowering their scores.
Our ablation (\S\ref{sec:rq2}), however, shows the best $\gamma$ is in $\{-0.5, -0.25, 0\}$ and \emph{never positive} across all three backbones and all five datasets.
The reason is that an interaction reveals \emph{topical attention} even when the experience was negative: a user who rated a horror movie 1 star still revealed a taste for horror, so the click signals topical interest while the rating signals a bad experience.
With $\gamma < 0$ the disliked item enters with a small \emph{positive} weight ($-\gamma > 0$, but below the $+1$ given to likes), so the filter still pushes recommendation mass toward spectrally similar items, just less aggressively.

\subsection{Component B: Signed Item-Item Operator}
\label{sec:method-laplacian}

The second component replaces the item-item operator $\mathbf{M}$ with a signed version that blends positive and negative co-interaction structure.
For Laplacian-based backbones (ChebyCF, GF-CF), we define a signed Laplacian
\begin{equation}
  \Lsign = \mathbf{I} - \Rtil^{+\top}\Rtil^{+} + \kappa\,\Rtil^{-\top}\Rtil^{-},
  \label{eq:signed-laplacian}
\end{equation}
where $\Rtil^{+}$ and $\Rtil^{-}$ are the normalized positive and negative interaction matrices and $\kappa \in \RR_{\geq 0}$ is a second hyperparameter; keeping $\kappa$ non-negative adds a positive penalty to item pairs that users frequently dislike together, so the backbone's low-pass filter downweights them in the predicted scores.\footnote{Equivalently, $\kappa \ge 0$ keeps $\kappa\,\Rtil^{-\top}\Rtil^{-}$ positive semi-definite, so the signed Laplacian has eigenvalues in $[0, 2]$~\citep{Singh2022SignedGraphNeural} and the low-pass filter stays well-posed.}
For Turbo-CF, which operates on the raw item-item similarity $\hat{\mathbf{P}}$ rather than a Laplacian, the natural analogue is a signed kernel $\hat{\mathbf{P}}^{\pm} = \Rtil^{+\top}\Rtil^{+} + \kappa\,\Rtil^{-\top}\Rtil^{-}$.

\paragraph{Why $\kappa \approx 0.1$ Suffices for Laplacian Backbones.}
Our joint sweep (\S\ref{sec:rq2}) shows $\kappa = 0.1$ optimal across all five datasets for \sysCheby; $\kappa > 0.1$ consistently hurts since the negative item-item graph is noisier than the positive one, so the model benefits from only a small perturbation of the positive spectrum, while on \sysGF the best $\kappa$ is $0$ on every dataset, so its gains come from component A alone (Table~\ref{tab:transfer-brief}).

A \system instance composes the two components on a chosen backbone: substitute $\mathbf{M}^{\pm}$ for $\mathbf{M}$ and $\mathbf{r}_u^{\pm}$ for $\mathbf{r}_u$ in $\hat{\mathbf{r}}_u = F(\mathbf{M})\mathbf{r}_u$.
We sweep a coarse grid, $\gamma \in \{-0.5, -0.25, 0, 0.25\}$ and $\kappa \in \{0, 0.1, 0.5, 1.0\}$, on top of per-dataset retuned backbone hyperparameters; the ranges follow the design of the two components: $\gamma$ brackets $0$ on both sides to test whether an explicit dislike acts as evidence for or against spectrally similar items, and $\kappa$ stays small and non-negative to keep the signed operator well-posed (\S\ref{sec:method-laplacian}).
The pipeline remains training-free: no gradient computation and no learned embeddings, and both components preserve the backbone's asymptotic cost.
The signed input $\mathbf{r}_u^{\pm}$ adds at most $\text{nnz}(\mathbf{r}_u^{+}) + \text{nnz}(\mathbf{r}_u^{-})$ non-zeros and the signed operator $\mathbf{M}^{\pm}$ shares the unsigned operator's sparsity pattern, requiring no eigendecomposition for the polynomial-filter backbones, so each instance inherits its backbone's $O(K \cdot \text{nnz}(\mathbf{R}))$ complexity for a degree-$K$ filter, where $K$ is the polynomial order (GF-CF is the $K{=}1$ case).

\section{Experiments}
\label{sec:experiment}
We organize the experiments around four research questions:

\begin{description}[nosep, leftmargin=!]
\item[RQ1 (Accuracy, \S\ref{sec:rq1})] Do the signed components improve recommendation quality over the unsigned backbone, and how do they compare to learned sign-aware models?
\item[RQ2 (Generality, \S\ref{sec:rq2})] Do the components transfer across heterogeneous spectral backbones?
\item[RQ3 (Efficiency, \S\ref{sec:rq3})] What is the cost of sign-awareness, and where does the family sit on the efficiency-accuracy Pareto frontier?
\item[RQ4 (Cold-start, \S\ref{sec:rq4})] When users have thin positive histories, how much does sign-awareness help?
\end{description}

\subsection{Setup}
\label{sec:datasets}

We evaluate on the five sign-aware benchmarks introduced by SIGformer~\citep{Chen2024SIGformerSignawareGraph}: Amazon-CDs and Amazon-Music~\citep{Ni2019JustifyingRecommendationsUsing}, Epinions~\citep{Tang2012ETrustUnderstandingTrust}, KuaiRand~\citep{Gao2022KuaiRandUnbiasedSequential}, and KuaiRec~\citep{Gao2022KuaiRecFullyobservedDataset}.
The five datasets span 1,411 to 51,267 users and 25,592 to 512,216 positive interactions, with positive-to-negative ratios from 1:0.22 (Amazon-CDs) to 1:5.95 (KuaiRec).
Negatives derive from 1-star reviews (Amazon-CDs, Amazon-Music), user-to-user distrust (Epinions), hate-button clicks (KuaiRand), and very-low watch ratios (KuaiRec).
We use the full-ranking top-$K$ protocol (each test user is ranked over all items not seen in training, with observed positives and negatives masked from candidates) and report Recall@20.
For each (dataset, backbone) pair, we first tune the backbone's native hyperparameters (e.g., the Chebyshev order for ChebyCF, the similarity exponent and polynomial degree for Turbo-CF, and the $\alpha$/$\eta$ weights for GF-CF) against Recall@20 on the unsigned positive-only interaction matrix, matching the original backbone setup, then apply $\gamma$ and $\kappa$ on top.
We compare \system against the unsigned spectral baselines Turbo-CF, GF-CF, BSPM~\citep{Choi2023BlurringSharpeningProcessModels} (bug-fixed), and ChebyCF, the learned graph baseline LightGCN, and the learned sign-aware baseline SIGformer.

\subsection{RQ1: Accuracy}
\label{sec:rq1}

\begin{table}[t]
  \centering
  \caption{Main results: Recall@20. \textbf{Bold}/\underline{underline} = best/second-best within the training-free block (unsigned + ours); learned models are shown for reference. The slowdown next to each learned method is its mean wall-clock relative to \sysCheby, averaged over the five datasets; per-dataset, SIGformer ranges from 7.7 to 155.3$\times$ slower (full per-dataset table omitted for space). Each \system instance matches or beats its unsigned backbone on all 5 datasets.}
  \label{tab:main-brief}
  \small
  \setlength{\tabcolsep}{2.6pt}
\begin{tabular}{lccccc}
  \toprule
  Method & A-CDs & A-Music & Epi. & K-Rand & K-Rec \\
  \midrule
  \multicolumn{6}{l}{\textit{Unsigned backbones (extended by \system)}} \\
  Turbo-CF & .1035 & .2495 & .0611 & .0905 & .0480 \\
  GF-CF    & .1136 & .2606 & \underline{.0815} & .1133 & .0438 \\
  ChebyCF  & .1122 & .2686 & .0694 & .1154 & \underline{.0630} \\
  \midrule
  \multicolumn{6}{l}{\textit{Other training-free baseline}} \\
  BSPM     & .1049 & .2558 & .0696 & .0987 & .0379 \\
  \midrule
  \multicolumn{6}{l}{\textit{Training-free, sign-aware (ours)}} \\
  \textbf{\sysTurbo} & .1105            & .2627            & .0698            & \textbf{.1213}   & .0533 \\
  \textbf{\sysGF}    & \textbf{.1183}   & \underline{.2706} & \textbf{.0848}  & \underline{.1212} & .0438 \\
  \textbf{\sysCheby} & \underline{.1173} & \textbf{.2775}   & .0805            & .1211            & \textbf{.0642} \\
  \midrule
  \multicolumn{6}{l}{\textit{Learned reference (gradient-based)}} \\
  \color{gray} LightGCN ($\sim$75.9$\times$ slower)  & \color{gray} .1333 & \color{gray} .2723 & \color{gray} .0755 & \color{gray} .0807 & \color{gray} .0425 \\
  \color{gray} SIGformer ($\sim$11.3$\times$ slower) & \color{gray} .1409 & \color{gray} .3059 & \color{gray} .0967 & \color{gray} .1487 & \color{gray} .0908 \\
  \bottomrule
\end{tabular}

\end{table}

Table~\ref{tab:main-brief} summarizes the accuracy comparison.
Each \system instance matches or beats its unsigned backbone on all 5 datasets: \sysCheby lifts ChebyCF by +1.9\% (KuaiRec) to +16.0\% (Epinions) R@20, \sysTurbo lifts Turbo-CF by +5.3\% (Amazon-Music) to +34.0\% (KuaiRand), and \sysGF lifts GF-CF by +3.8\% (Amazon-Music) to +7.0\% (KuaiRand) on 4 datasets with a tie on KuaiRec; \sysGF achieves the best Epinions R@20 (0.0848) among all training-free methods, and \sysCheby beats LightGCN on 4 of 5 datasets despite being training-free.

\takeaway{Every \system instance matches or beats its unsigned backbone on all 5 sign-aware benchmarks, and \sysCheby beats LightGCN on 4 of 5 without any gradient updates.}

\subsection{RQ2: Generality Across Spectral Backbones}
\label{sec:rq2}

\begin{table}[t]
  \centering
  \caption{Transfer across three spectral backbones: $\Delta$ Recall@20 over the unsigned backbone at the best $(\gamma,\kappa)$ for each cell, which need not match across cells; the full $4{\times}4$ sweep is omitted for space.}
  \label{tab:transfer-brief}
  \small
  \setlength{\tabcolsep}{3.0pt}
\begin{tabular}{lrrrrr}
  \toprule
  Backbone & A-CDs & A-Music & Epi. & K-Rand & K-Rec \\
  \midrule
  GF-CF    & +3.5\% & +3.2\%  & +3.4\%  & +5.1\%  & \phantom{+}0.0\% \\
  Turbo-CF & +6.3\% & +4.5\%  & +13.2\% & +32.6\% & +10.3\% \\
  ChebyCF  & +4.5\% & +3.3\%  & +15.9\% & +5.0\%  & +1.9\% \\
  \bottomrule
\end{tabular}

\end{table}

Table~\ref{tab:transfer-brief} reports the Recall@20 lift at the best $(\gamma,\kappa)$ for each (dataset, backbone) pair; this best setting can differ from cell to cell, and the full $4{\times}4$ sweep behind it is omitted for space.
Three findings hold across all 15 cells.
First, the best $\gamma$ is \emph{always} in $\{-0.5, -0.25, 0\}$ and never positive, supporting the attention interpretation in \S\ref{sec:method-signal} as a backbone-agnostic property.
Second, the default $(\gamma{=}{-}0.5,\,\kappa{=}0.1)$ suffices for \sysCheby on 4 of 5 datasets; \sysGF and \sysTurbo benefit from per-dataset $\kappa$ but from no other tuning.
Third, Turbo-CF benefits the most (+4.5\% to +32.6\%, median +10.3\%) versus ChebyCF (median +4.5\%) and GF-CF (median +3.4\%): its simpler polynomial filter leaves more headroom for the negative signal to add information.

\takeaway{The components transfer: Recall@20 lifts reach +32.6\% with backbone-specific tuning and +1.9\% to +16.0\% for \sysCheby at the fixed default $(\gamma{=}{-}0.5,\kappa{=}0.1)$; the best $\gamma$ is non-positive in all 15 cells, and the default holds for \sysCheby on 4 of 5 datasets.}

\subsection{RQ3: Efficiency and Pareto Frontier}
\label{sec:rq3}

\begin{table}[t]
  \centering
  \caption{Efficiency summary (means across the five datasets). \emph{Overhead}: time added over the unsigned backbone; \emph{R@20\,/\,SIG}: Recall@20 normalized to SIGformer.}
  \label{tab:efficiency-brief}
  \small
  \setlength{\tabcolsep}{5pt}
\begin{tabular}{lrrr}
  \toprule
  Method & Avg time (s) & Overhead & R@20 / SIG \\
  \midrule
  \sysGF    & 8.5  & +25.0\% & 0.780 \\
  \sysTurbo & 16.2 & +58.8\% & 0.754 \\
  \sysCheby & 23.8 & +16.1\% & \textbf{0.819} \\
  \midrule
  \color{gray} SIGformer & \color{gray} 270.0 & \color{gray} n/a & \color{gray} 1.000 \\
  \bottomrule
\end{tabular}

\end{table}

Figure~\ref{fig:pipeline}(d) shows the efficiency-accuracy tradeoff and Table~\ref{tab:efficiency-brief} the per-instance cost.
Adding sign-awareness is cheap: each \system instance adds only 16.1\% to 58.8\% wall-clock over its unsigned backbone.
The cheapest signed instance, \sysGF (8.5\,s on average), dominates Turbo-CF and BSPM on both axes.
Against the learned SIGformer, the family runs 7.7 to 155.3$\times$ faster while trailing its Recall@20 by only 10.2\% to 41.4\% per dataset.

\takeaway{Sign-awareness costs 16.1\% to 58.8\% extra wall-clock; \sysGF and \sysCheby sit on the training-free Pareto frontier, and the family runs 7.7 to 155.3$\times$ faster than SIGformer at 70.7\% to 90.7\% of its R@20.}

\subsection{RQ4: Cold-Start Users}
\label{sec:rq4}

\begin{table}[t]
  \centering
  \caption{Cold-start lift: $\Delta$ Recall@20 of \sysCheby over ChebyCF on users with 1 to 5 training items (the $[1, 6)$ bucket).}
  \label{tab:coldstart-brief}
  \small
  \setlength{\tabcolsep}{5pt}
\begin{tabular}{lccccc}
  \toprule
              & A-CDs    & A-Music  & Epi.              & K-Rand   & K-Rec    \\
  \midrule
  $\Delta$ R@20 & +5.8\%  & +7.0\%   & \textbf{+29.2\%}  & +3.5\%   & +2.2\%   \\
  \bottomrule
\end{tabular}

\end{table}

On cold-start users with 1 to 5 training items, \sysCheby lifts Recall@20 over ChebyCF on every dataset: +29.2\% (Epinions), +7.0\% (Amazon-Music), +5.8\% (Amazon-CDs), +3.5\% (KuaiRand), and +2.2\% (KuaiRec).
Table~\ref{tab:coldstart-brief} reports only this most-favorable, cold-start bucket (full per-bucket breakdown omitted for space); across buckets the lift shrinks as positive histories grow and turns substantially negative for the most active users (up to -11.1\% on KuaiRec at 100 or more items), motivating a per-user adaptive $\gamma$ as future work.

\takeaway{Sign-awareness helps most for cold-start users (up to +29.2\% R@20 on Epinions) but the lift can turn negative for active users on noisier datasets.}

\section{Conclusion}
\label{sec:conclusion}

\system extends any spectral CF backbone with two training-free components $(\mathbf{r}_u^{\pm}, \mathbf{M}^{\pm})$; on ChebyCF, GF-CF, and Turbo-CF it matches or beats its unsigned backbone on all 5 benchmarks at 7.7 to 155.3$\times$ SIGformer speedup and 70.7\% to 90.7\% R@20.
Two of \system's design choices each invite a natural extension. First, its single global $\gamma$ over-penalizes active users whose likes and dislikes span different topics, motivating a per-user adaptive $\gamma$. Second, its fixed linear filter could give way to richer backbones such as BSPM's blurring-sharpening process and PolyCF's learnable filters.

\newpage
\section*{GenAI Usage Disclosure}

To assist with technical implementation, the authors used Claude Code and the Gemini CLI for creating plots and diagrams. 
All core intellectual contributions, including the algorithm design, experimental methodology, and original manuscript, were developed solely by the authors. 
The use of LLMs was strictly confined to these technical tasks and refining the readability of the authors' text. 
AI was not employed to conceive research concepts, produce data, or conduct evaluations.

\bibliographystyle{ACM-Reference-Format}
\bibliography{zotero}


\begin{thebibliography}{28}


\ifx \showCODEN    \undefined \def \showCODEN     #1{\unskip}     \fi
\ifx \showISBNx    \undefined \def \showISBNx     #1{\unskip}     \fi
\ifx \showISBNxiii \undefined \def \showISBNxiii  #1{\unskip}     \fi
\ifx \showISSN     \undefined \def \showISSN      #1{\unskip}     \fi
\ifx \showLCCN     \undefined \def \showLCCN      #1{\unskip}     \fi
\ifx \shownote     \undefined \def \shownote      #1{#1}          \fi
\ifx \showarticletitle \undefined \def \showarticletitle #1{#1}   \fi
\ifx \showURL      \undefined \def \showURL       {\relax}        \fi
\providecommand\bibfield[2]{#2}
\providecommand\bibinfo[2]{#2}
\providecommand\natexlab[1]{#1}
\providecommand\showeprint[2][]{arXiv:#2}

\bibitem[Cai et~al\mbox{.}(2023)]%
        {Cai2023LightGCLSimpleEffective}
\bibfield{author}{\bibinfo{person}{Xuheng Cai}, \bibinfo{person}{Chao Huang},
  \bibinfo{person}{Lianghao Xia}, {and} \bibinfo{person}{Xubin Ren}.}
  \bibinfo{year}{2023}\natexlab{}.
\newblock \bibinfo{title}{{{LightGCL}}: {{Simple Yet Effective Graph
  Contrastive Learning}} for {{Recommendation}}}.
\newblock
\showeprint[arxiv]{2302.08191}~[cs]
\href{https://doi.org/10.48550/arXiv.2302.08191}{doi:\nolinkurl{10.48550/arXiv.2302.08191}}


\bibitem[Chen et~al\mbox{.}(2024)]%
        {Chen2024SIGformerSignawareGraph}
\bibfield{author}{\bibinfo{person}{Sirui Chen}, \bibinfo{person}{Jiawei Chen},
  \bibinfo{person}{Sheng Zhou}, \bibinfo{person}{Bohao Wang},
  \bibinfo{person}{Shen Han}, \bibinfo{person}{Chanfei Su},
  \bibinfo{person}{Yuqing Yuan}, {and} \bibinfo{person}{Can Wang}.}
  \bibinfo{year}{2024}\natexlab{}.
\newblock \showarticletitle{{{SIGformer}}: {{Sign-aware Graph Transformer}} for
  {{Recommendation}}}. In \bibinfo{booktitle}{\emph{Proceedings of the 47th
  {{International ACM SIGIR Conference}} on {{Research}} and {{Development}} in
  {{Information Retrieval}}}}. \bibinfo{pages}{1274--1284}.
\newblock
\showeprint[arxiv]{2404.11982}~[cs]
\href{https://doi.org/10.1145/3626772.3657747}{doi:\nolinkurl{10.1145/3626772.3657747}}


\bibitem[Choi et~al\mbox{.}(2023)]%
        {Choi2023BlurringSharpeningProcessModels}
\bibfield{author}{\bibinfo{person}{Jeongwhan Choi}, \bibinfo{person}{Seoyoung
  Hong}, \bibinfo{person}{Noseong Park}, {and} \bibinfo{person}{Sung-Bae Cho}.}
  \bibinfo{year}{2023}\natexlab{}.
\newblock \bibinfo{title}{Blurring-{{Sharpening Process Models}} for
  {{Collaborative Filtering}}}.
\newblock
\showeprint[arxiv]{2211.09324}~[cs]
\href{https://doi.org/10.48550/arXiv.2211.09324}{doi:\nolinkurl{10.48550/arXiv.2211.09324}}


\bibitem[D'Amico et~al\mbox{.}(2023)]%
        {DAmico2023PureSpectralGraph}
\bibfield{author}{\bibinfo{person}{Edoardo D'Amico}, \bibinfo{person}{Aonghus
  Lawlor}, {and} \bibinfo{person}{Neil Hurley}.}
  \bibinfo{year}{2023}\natexlab{}.
\newblock \showarticletitle{Pure {{Spectral Graph Embeddings}}:
  {{Reinterpreting Graph Convolution}} for {{Top-N Recommendation}}}.
\newblock In \bibinfo{booktitle}{\emph{Advances in {{Knowledge Discovery}} and
  {{Data Mining}}}}, \bibfield{editor}{\bibinfo{person}{Hisashi Kashima},
  \bibinfo{person}{Tsuyoshi Ide}, {and} \bibinfo{person}{Wen-Chih Peng}}
  (Eds.). Vol.~\bibinfo{volume}{13937}. \bibinfo{publisher}{Springer Nature
  Switzerland}, \bibinfo{address}{Cham}, \bibinfo{pages}{310--321}.
\newblock
\showISBNx{978-3-031-33379-8 978-3-031-33380-4}
\href{https://doi.org/10.1007/978-3-031-33380-4_24}{doi:\nolinkurl{10.1007/978-3-031-33380-4_24}}


\bibitem[Gao et~al\mbox{.}(2022a)]%
        {Gao2022KuaiRecFullyobservedDataset}
\bibfield{author}{\bibinfo{person}{Chongming Gao}, \bibinfo{person}{Shijun Li},
  \bibinfo{person}{Wenqiang Lei}, \bibinfo{person}{Jiawei Chen},
  \bibinfo{person}{Biao Li}, \bibinfo{person}{Peng Jiang},
  \bibinfo{person}{Xiangnan He}, \bibinfo{person}{Jiaxin Mao}, {and}
  \bibinfo{person}{Tat-Seng Chua}.} \bibinfo{year}{2022}\natexlab{a}.
\newblock \showarticletitle{{{KuaiRec}}: {{A Fully-observed Dataset}} and
  {{Insights}} for {{Evaluating Recommender Systems}}}. In
  \bibinfo{booktitle}{\emph{Proceedings of the 31st {{ACM International
  Conference}} on {{Information}} \& {{Knowledge Management}}}}.
  \bibinfo{pages}{540--550}.
\newblock
\showeprint[arxiv]{2202.10842}~[cs]
\href{https://doi.org/10.1145/3511808.3557220}{doi:\nolinkurl{10.1145/3511808.3557220}}


\bibitem[Gao et~al\mbox{.}(2022b)]%
        {Gao2022KuaiRandUnbiasedSequential}
\bibfield{author}{\bibinfo{person}{Chongming Gao}, \bibinfo{person}{Shijun Li},
  \bibinfo{person}{Yuan Zhang}, \bibinfo{person}{Jiawei Chen},
  \bibinfo{person}{Biao Li}, \bibinfo{person}{Wenqiang Lei},
  \bibinfo{person}{Peng Jiang}, {and} \bibinfo{person}{Xiangnan He}.}
  \bibinfo{year}{2022}\natexlab{b}.
\newblock \showarticletitle{{{KuaiRand}}: {{An Unbiased Sequential
  Recommendation Dataset}} with {{Randomly Exposed Videos}}}. In
  \bibinfo{booktitle}{\emph{Proceedings of the 31st {{ACM International
  Conference}} on {{Information}} \& {{Knowledge Management}}}}.
  \bibinfo{pages}{3953--3957}.
\newblock
\showeprint[arxiv]{2208.08696}~[cs]
\href{https://doi.org/10.1145/3511808.3557624}{doi:\nolinkurl{10.1145/3511808.3557624}}


\bibitem[Guo et~al\mbox{.}(2023)]%
        {Guo2023ManipulatingSignalsUserItem}
\bibfield{author}{\bibinfo{person}{Jiayan Guo}, \bibinfo{person}{Lun Du},
  \bibinfo{person}{Xu Chen}, \bibinfo{person}{Xiaojun Ma},
  \bibinfo{person}{Qiang Fu}, \bibinfo{person}{Shi Han},
  \bibinfo{person}{Dongmei Zhang}, {and} \bibinfo{person}{Yan Zhang}.}
  \bibinfo{year}{2023}\natexlab{}.
\newblock \showarticletitle{On {{Manipulating Signals}} of {{User-Item Graph}}:
  {{A Jacobi Polynomial-based Graph Collaborative Filtering}}}. In
  \bibinfo{booktitle}{\emph{Proceedings of the 29th {{ACM SIGKDD Conference}}
  on {{Knowledge Discovery}} and {{Data Mining}}}}. \bibinfo{publisher}{ACM},
  \bibinfo{address}{Long Beach CA USA}, \bibinfo{pages}{602--613}.
\newblock
\showISBNx{979-8-4007-0103-0}
\href{https://doi.org/10.1145/3580305.3599450}{doi:\nolinkurl{10.1145/3580305.3599450}}


\bibitem[He et~al\mbox{.}(2020)]%
        {He2020LightGCNSimplifyingPowering}
\bibfield{author}{\bibinfo{person}{Xiangnan He}, \bibinfo{person}{Kuan Deng},
  \bibinfo{person}{Xiang Wang}, \bibinfo{person}{Yan Li},
  \bibinfo{person}{Yongdong Zhang}, {and} \bibinfo{person}{Meng Wang}.}
  \bibinfo{year}{2020}\natexlab{}.
\newblock \bibinfo{title}{{{LightGCN}}: {{Simplifying}} and {{Powering Graph
  Convolution Network}} for {{Recommendation}}}.
\newblock
\showeprint[arxiv]{2002.02126}~[cs]
\href{https://doi.org/10.48550/arXiv.2002.02126}{doi:\nolinkurl{10.48550/arXiv.2002.02126}}


\bibitem[Kim et~al\mbox{.}(2025)]%
        {Kim2025GraphSpectralFiltering}
\bibfield{author}{\bibinfo{person}{Chanwoo Kim}, \bibinfo{person}{Jinkyu Sung},
  \bibinfo{person}{Yebonn Han}, {and} \bibinfo{person}{Joonseok Lee}.}
  \bibinfo{year}{2025}\natexlab{}.
\newblock \bibinfo{title}{Graph {{Spectral Filtering}} with {{Chebyshev
  Interpolation}} for {{Recommendation}}}.
\newblock
\showeprint[arxiv]{2505.00552}~[cs]
\href{https://doi.org/10.48550/arXiv.2505.00552}{doi:\nolinkurl{10.48550/arXiv.2505.00552}}


\bibitem[Liu et~al\mbox{.}(2026)]%
        {Liu2026UnifiedModelingPositive}
\bibfield{author}{\bibinfo{person}{Yuting Liu}, \bibinfo{person}{Yizhou Dang},
  \bibinfo{person}{Yuliang Liang}, \bibinfo{person}{Qiang Liu},
  \bibinfo{person}{Guibing Guo}, \bibinfo{person}{Jianzhe Zhao}, {and}
  \bibinfo{person}{Xingwei Wang}.} \bibinfo{year}{2026}\natexlab{}.
\newblock \showarticletitle{Towards {{Unified Modeling}} for {{Positive}} and
  {{Negative Preferences}} in {{Sign-Aware Recommendation}}}.
\newblock In \bibinfo{booktitle}{\emph{Database {{Systems}} for {{Advanced
  Applications}}}}, \bibfield{editor}{\bibinfo{person}{Feida Zhu},
  \bibinfo{person}{Philip~S. Yu}, \bibinfo{person}{Akiyo Nadamoto},
  \bibinfo{person}{Ee-peng Lim}, \bibinfo{person}{Kyuseok Shim},
  \bibinfo{person}{Wei Ding}, {and} \bibinfo{person}{Bingxue Zhang}} (Eds.).
  Vol.~\bibinfo{volume}{15990}. \bibinfo{publisher}{Springer Nature Singapore},
  \bibinfo{address}{Singapore}, \bibinfo{pages}{303--319}.
\newblock
\showISBNx{9789819541546 9789819541553}
\href{https://doi.org/10.1007/978-981-95-4155-3_19}{doi:\nolinkurl{10.1007/978-981-95-4155-3_19}}


\bibitem[Liu et~al\mbox{.}(2023)]%
        {Liu2023PANEGNNUnifyingPositive}
\bibfield{author}{\bibinfo{person}{Ziyang Liu}, \bibinfo{person}{Chaokun Wang},
  \bibinfo{person}{Jingcao Xu}, \bibinfo{person}{Cheng Wu},
  \bibinfo{person}{Kai Zheng}, \bibinfo{person}{Yang Song}, \bibinfo{person}{Na
  Mou}, {and} \bibinfo{person}{Kun Gai}.} \bibinfo{year}{2023}\natexlab{}.
\newblock \bibinfo{title}{{{PANE-GNN}}: {{Unifying Positive}} and {{Negative
  Edges}} in {{Graph Neural Networks}} for {{Recommendation}}}.
\newblock
\showeprint[arxiv]{2306.04095}~[cs]
\href{https://doi.org/10.48550/arXiv.2306.04095}{doi:\nolinkurl{10.48550/arXiv.2306.04095}}


\bibitem[Ni et~al\mbox{.}(2019)]%
        {Ni2019JustifyingRecommendationsUsing}
\bibfield{author}{\bibinfo{person}{Jianmo Ni}, \bibinfo{person}{Jiacheng Li},
  {and} \bibinfo{person}{Julian McAuley}.} \bibinfo{year}{2019}\natexlab{}.
\newblock \showarticletitle{Justifying {{Recommendations}} Using
  {{Distantly-Labeled Reviews}} and {{Fine-Grained Aspects}}}. In
  \bibinfo{booktitle}{\emph{Proceedings of the 2019 {{Conference}} on
  {{Empirical Methods}} in {{Natural Language Processing}} and the 9th
  {{International Joint Conference}} on {{Natural Language Processing}}
  ({{EMNLP-IJCNLP}})}}, \bibfield{editor}{\bibinfo{person}{Kentaro Inui},
  \bibinfo{person}{Jing Jiang}, \bibinfo{person}{Vincent Ng}, {and}
  \bibinfo{person}{Xiaojun Wan}} (Eds.). \bibinfo{publisher}{Association for
  Computational Linguistics}, \bibinfo{address}{Hong Kong, China},
  \bibinfo{pages}{188--197}.
\newblock
\href{https://doi.org/10.18653/v1/D19-1018}{doi:\nolinkurl{10.18653/v1/D19-1018}}


\bibitem[Park et~al\mbox{.}(2024)]%
        {Park2024TurboCFMatrixDecompositionFree}
\bibfield{author}{\bibinfo{person}{Jin-Duk Park}, \bibinfo{person}{Yong-Min
  Shin}, {and} \bibinfo{person}{Won-Yong Shin}.}
  \bibinfo{year}{2024}\natexlab{}.
\newblock \bibinfo{title}{Turbo-{{CF}}: {{Matrix Decomposition-Free Graph
  Filtering}} for {{Fast Recommendation}}}.
\newblock
\showeprint[arxiv]{2404.14243}~[cs]
\href{https://doi.org/10.48550/arXiv.2404.14243}{doi:\nolinkurl{10.48550/arXiv.2404.14243}}


\bibitem[Peng et~al\mbox{.}(2022)]%
        {Peng2022LessMoreReweighting}
\bibfield{author}{\bibinfo{person}{Shaowen Peng}, \bibinfo{person}{Kazunari
  Sugiyama}, {and} \bibinfo{person}{Tsunenori Mine}.}
  \bibinfo{year}{2022}\natexlab{}.
\newblock \showarticletitle{Less Is {{More}}: {{Reweighting Important Spectral
  Graph Features}} for {{Recommendation}}}. In
  \bibinfo{booktitle}{\emph{Proceedings of the 45th {{International ACM SIGIR
  Conference}} on {{Research}} and {{Development}} in {{Information
  Retrieval}}}}. \bibinfo{publisher}{ACM}, \bibinfo{address}{Madrid Spain},
  \bibinfo{pages}{1273--1282}.
\newblock
\showISBNx{978-1-4503-8732-3}
\href{https://doi.org/10.1145/3477495.3532014}{doi:\nolinkurl{10.1145/3477495.3532014}}


\bibitem[Qin et~al\mbox{.}(2024)]%
        {Qin2024PolyCFOptimalSpectral}
\bibfield{author}{\bibinfo{person}{Yifang Qin}, \bibinfo{person}{Wei Ju},
  \bibinfo{person}{Xiao Luo}, \bibinfo{person}{Yiyang Gu},
  \bibinfo{person}{Zhiping Xiao}, {and} \bibinfo{person}{Ming Zhang}.}
  \bibinfo{year}{2024}\natexlab{}.
\newblock \bibinfo{title}{{{PolyCF}}: {{Towards}} the {{Optimal Spectral Graph
  Filters}} for {{Collaborative Filtering}}}.
\newblock
\showeprint[arxiv]{2401.12590}~[cs]
\href{https://doi.org/10.48550/arXiv.2401.12590}{doi:\nolinkurl{10.48550/arXiv.2401.12590}}


\bibitem[Seo et~al\mbox{.}(2022)]%
        {Seo2022SiReNSignAwareRecommendation}
\bibfield{author}{\bibinfo{person}{Changwon Seo}, \bibinfo{person}{Kyeong-Joong
  Jeong}, \bibinfo{person}{Sungsu Lim}, {and} \bibinfo{person}{Won-Yong Shin}.}
  \bibinfo{year}{2022}\natexlab{}.
\newblock \bibinfo{title}{{{SiReN}}: {{Sign-Aware Recommendation Using Graph
  Neural Networks}}}.
\newblock
\showeprint[arxiv]{2108.08735}~[cs]
\href{https://doi.org/10.48550/arXiv.2108.08735}{doi:\nolinkurl{10.48550/arXiv.2108.08735}}


\bibitem[Shen et~al\mbox{.}(2021)]%
        {Shen2021HowPowerfulGraph}
\bibfield{author}{\bibinfo{person}{Yifei Shen}, \bibinfo{person}{Yongji Wu},
  \bibinfo{person}{Yao Zhang}, \bibinfo{person}{Caihua Shan},
  \bibinfo{person}{Jun Zhang}, \bibinfo{person}{Khaled~B. Letaief}, {and}
  \bibinfo{person}{Dongsheng Li}.} \bibinfo{year}{2021}\natexlab{}.
\newblock \bibinfo{title}{How {{Powerful}} Is {{Graph Convolution}} for
  {{Recommendation}}?}
\newblock
\showeprint[arxiv]{2108.07567}~[cs]
\href{https://doi.org/10.48550/arXiv.2108.07567}{doi:\nolinkurl{10.48550/arXiv.2108.07567}}


\bibitem[Singh and Chen(2022)]%
        {Singh2022SignedGraphNeural}
\bibfield{author}{\bibinfo{person}{Rahul Singh} {and} \bibinfo{person}{Yongxin
  Chen}.} \bibinfo{year}{2022}\natexlab{}.
\newblock \bibinfo{title}{Signed {{Graph Neural Networks}}: {{A Frequency
  Perspective}}}.
\newblock
\showeprint[arxiv]{2208.07323}~[cs]
\href{https://doi.org/10.48550/arXiv.2208.07323}{doi:\nolinkurl{10.48550/arXiv.2208.07323}}


\bibitem[Steck(2019)]%
        {Steck2019EmbarrassinglyShallowAutoencoders}
\bibfield{author}{\bibinfo{person}{Harald Steck}.}
  \bibinfo{year}{2019}\natexlab{}.
\newblock \showarticletitle{Embarrassingly {{Shallow Autoencoders}} for
  {{Sparse Data}}}. In \bibinfo{booktitle}{\emph{The {{World Wide Web
  Conference}}}}. \bibinfo{publisher}{ACM}, \bibinfo{address}{San Francisco CA
  USA}, \bibinfo{pages}{3251--3257}.
\newblock
\showISBNx{978-1-4503-6674-8}
\href{https://doi.org/10.1145/3308558.3313710}{doi:\nolinkurl{10.1145/3308558.3313710}}


\bibitem[Tang et~al\mbox{.}(2012)]%
        {Tang2012ETrustUnderstandingTrust}
\bibfield{author}{\bibinfo{person}{Jiliang Tang}, \bibinfo{person}{Huiji Gao},
  \bibinfo{person}{Huan Liu}, {and} \bibinfo{person}{Atish Das~Sarma}.}
  \bibinfo{year}{2012}\natexlab{}.
\newblock \showarticletitle{{{eTrust}}: {{Understanding Trust Evolution}} in an
  {{Online World}}}. In \bibinfo{booktitle}{\emph{Proceedings of the 18th {{ACM
  SIGKDD}} International Conference on {{Knowledge}} Discovery and Data
  Mining}} \emph{(\bibinfo{series}{{{KDD}} '12})}.
  \bibinfo{publisher}{Association for Computing Machinery},
  \bibinfo{address}{New York, NY, USA}, \bibinfo{pages}{253--261}.
\newblock
\showISBNx{978-1-4503-1462-6}
\href{https://doi.org/10.1145/2339530.2339574}{doi:\nolinkurl{10.1145/2339530.2339574}}


\bibitem[Wang et~al\mbox{.}(2025)]%
        {Wang2025GegenNetSpectralConvolutional}
\bibfield{author}{\bibinfo{person}{Hewen Wang}, \bibinfo{person}{Renchi Yang},
  {and} \bibinfo{person}{Xiaokui Xiao}.} \bibinfo{year}{2025}\natexlab{}.
\newblock \showarticletitle{{{GegenNet}}: {{Spectral Convolutional Neural
  Networks}} for {{Link Sign Prediction}} in {{Signed Bipartite Graphs}}}. In
  \bibinfo{booktitle}{\emph{Proceedings of the 34th {{ACM International
  Conference}} on {{Information}} and {{Knowledge Management}}}}.
  \bibinfo{publisher}{ACM}, \bibinfo{address}{Seoul Republic of Korea},
  \bibinfo{pages}{2987--2997}.
\newblock
\showISBNx{979-8-4007-2040-6}
\href{https://doi.org/10.1145/3746252.3761129}{doi:\nolinkurl{10.1145/3746252.3761129}}


\bibitem[Wang et~al\mbox{.}(2020)]%
        {Wang2020GraphLearningApproaches}
\bibfield{author}{\bibinfo{person}{Shoujin Wang}, \bibinfo{person}{Liang Hu},
  \bibinfo{person}{Yan Wang}, \bibinfo{person}{Xiangnan He},
  \bibinfo{person}{Quan~Z. Sheng}, \bibinfo{person}{Mehmet Orgun},
  \bibinfo{person}{Longbing Cao}, \bibinfo{person}{Nan Wang},
  \bibinfo{person}{Francesco Ricci}, {and} \bibinfo{person}{Philip~S. Yu}.}
  \bibinfo{year}{2020}\natexlab{}.
\newblock \bibinfo{title}{Graph {{Learning Approaches}} to {{Recommender
  Systems}}: {{A Review}}}.
\newblock
\showeprint[arxiv]{2004.11718}~[cs]
\href{https://doi.org/10.48550/arXiv.2004.11718}{doi:\nolinkurl{10.48550/arXiv.2004.11718}}


\bibitem[Wang et~al\mbox{.}(2024)]%
        {Wang2024NFARecNegativeFeedbackAware}
\bibfield{author}{\bibinfo{person}{Xinfeng Wang}, \bibinfo{person}{Fumiyo
  Fukumoto}, \bibinfo{person}{Jin Cui}, \bibinfo{person}{Yoshimi Suzuki}, {and}
  \bibinfo{person}{Dongjin Yu}.} \bibinfo{year}{2024}\natexlab{}.
\newblock \bibinfo{title}{{{NFARec}}: {{A Negative Feedback-Aware Recommender
  Model}}}.
\newblock
\showeprint[arxiv]{2404.06900}~[cs]
\href{https://doi.org/10.48550/arXiv.2404.06900}{doi:\nolinkurl{10.48550/arXiv.2404.06900}}


\bibitem[Wang et~al\mbox{.}(2019)]%
        {Wang2019NeuralGraphCollaborative}
\bibfield{author}{\bibinfo{person}{Xiang Wang}, \bibinfo{person}{Xiangnan He},
  \bibinfo{person}{Meng Wang}, \bibinfo{person}{Fuli Feng}, {and}
  \bibinfo{person}{Tat-Seng Chua}.} \bibinfo{year}{2019}\natexlab{}.
\newblock \showarticletitle{Neural {{Graph Collaborative Filtering}}}. In
  \bibinfo{booktitle}{\emph{Proceedings of the 42nd {{International ACM SIGIR
  Conference}} on {{Research}} and {{Development}} in {{Information
  Retrieval}}}}. \bibinfo{pages}{165--174}.
\newblock
\showeprint[arxiv]{1905.08108}~[cs]
\href{https://doi.org/10.1145/3331184.3331267}{doi:\nolinkurl{10.1145/3331184.3331267}}


\bibitem[Wu et~al\mbox{.}(2022)]%
        {Wu2022GraphNeuralNetworks}
\bibfield{author}{\bibinfo{person}{Shiwen Wu}, \bibinfo{person}{Fei Sun},
  \bibinfo{person}{Wentao Zhang}, \bibinfo{person}{Xu Xie}, {and}
  \bibinfo{person}{Bin Cui}.} \bibinfo{year}{2022}\natexlab{}.
\newblock \bibinfo{title}{Graph {{Neural Networks}} in {{Recommender Systems}}:
  {{A Survey}}}.
\newblock
\showeprint[arxiv]{2011.02260}~[cs]
\href{https://doi.org/10.48550/arXiv.2011.02260}{doi:\nolinkurl{10.48550/arXiv.2011.02260}}


\bibitem[Wu et~al\mbox{.}(2024)]%
        {Wu2024DFGNNDualfrequencyGraph}
\bibfield{author}{\bibinfo{person}{Yiqing Wu}, \bibinfo{person}{Ruobing Xie},
  \bibinfo{person}{Zhao Zhang}, \bibinfo{person}{Xu Zhang},
  \bibinfo{person}{Fuzhen Zhuang}, \bibinfo{person}{Leyu Lin},
  \bibinfo{person}{Zhanhui Kang}, {and} \bibinfo{person}{Yongjun Xu}.}
  \bibinfo{year}{2024}\natexlab{}.
\newblock \showarticletitle{{{DFGNN}}: {{Dual-frequency Graph Neural Network}}
  for {{Sign-aware Feedback}}}. In \bibinfo{booktitle}{\emph{Proceedings of the
  30th {{ACM SIGKDD Conference}} on {{Knowledge Discovery}} and {{Data
  Mining}}}}. \bibinfo{publisher}{ACM}, \bibinfo{address}{Barcelona Spain},
  \bibinfo{pages}{3437--3447}.
\newblock
\showISBNx{979-8-4007-0490-1}
\href{https://doi.org/10.1145/3637528.3671701}{doi:\nolinkurl{10.1145/3637528.3671701}}


\bibitem[Zhang et~al\mbox{.}(2023)]%
        {Zhang2023RecommendingGraphsComprehensive}
\bibfield{author}{\bibinfo{person}{Lemei Zhang}, \bibinfo{person}{Peng Liu},
  {and} \bibinfo{person}{Jon~Atle Gulla}.} \bibinfo{year}{2023}\natexlab{}.
\newblock \showarticletitle{Recommending on {{Graphs}}: {{A Comprehensive
  Review}} from a {{Data Perspective}}}.
\newblock \bibinfo{journal}{\emph{User Modeling and User-Adapted Interaction}}
  \bibinfo{volume}{33}, \bibinfo{number}{4} (\bibinfo{date}{Sept.}
  \bibinfo{year}{2023}), \bibinfo{pages}{803--888}.
\newblock
\showISSN{0924-1868, 1573-1391}
\showeprint[arxiv]{2212.12230}~[cs]
\href{https://doi.org/10.1007/s11257-023-09359-w}{doi:\nolinkurl{10.1007/s11257-023-09359-w}}


\bibitem[Zheng et~al\mbox{.}(2018)]%
        {Zheng2018SpectralCollaborativeFiltering}
\bibfield{author}{\bibinfo{person}{Lei Zheng}, \bibinfo{person}{Chun-Ta Lu},
  \bibinfo{person}{Fei Jiang}, \bibinfo{person}{Jiawei Zhang}, {and}
  \bibinfo{person}{Philip~S. Yu}.} \bibinfo{year}{2018}\natexlab{}.
\newblock \showarticletitle{Spectral Collaborative Filtering}. In
  \bibinfo{booktitle}{\emph{Proceedings of the 12th {{ACM Conference}} on
  {{Recommender Systems}}}}. \bibinfo{publisher}{ACM},
  \bibinfo{address}{Vancouver British Columbia Canada},
  \bibinfo{pages}{311--319}.
\newblock
\showISBNx{978-1-4503-5901-6}
\href{https://doi.org/10.1145/3240323.3240343}{doi:\nolinkurl{10.1145/3240323.3240343}}


\end{thebibliography}

\appendix
\section{Limitations and Future Work}
\label{sec:app-limitations}

\paragraph{Residual accuracy gap.}
\system recovers $70$--$91\%$ of SIGformer's Recall@20 across the five datasets, leaving a consistent $9$--$30\%$ gap.
Two factors drive this gap.
First, SIGformer's Transformer learns per-user, per-item weighting of the positive and negative graphs, whereas \system uses a single pair of scalars $(\gamma, \kappa)$ for all users on a dataset.
Second, the learned signed Laplacian in SIGformer effectively performs its own low-frequency decomposition during training; our closed-form filters inherit whatever low-pass shape the underlying backbone defines and cannot adapt the shape to the signed operator.

\paragraph{Noisier negative feedback.}
The joint $(\gamma, \kappa)$ sweep in \S\ref{sec:rq2} reveals one dataset, KuaiRec, on which the Laplacian backbones (\sysCheby, \sysGF) prefer $\gamma{=}0$ (ignore the negative signal entirely) and do not benefit from $\kappa > 0.1$: for GF-CF $\kappa$ has no measurable effect, and for ChebyCF $\kappa > 0.1$ hurts.
KuaiRec's negatives are derived from a watch-ratio threshold, which is a softer signal than explicit 1-star ratings or hate-button clicks.
This suggests a per-dataset annealing schedule on $\gamma$ tied to estimated negative-signal quality (for example, the calibration of raters who disagree on the same item) could recover some of the missing lift on datasets with weak negatives.

\paragraph{Future directions.}
Three extensions follow from the current design.
(1) \textbf{Per-user adaptive $\gamma$.}
Users whose positives and negatives form topically incoherent sets (e.g., they click movies across many genres but 1-star only horror) benefit from attenuating the negative channel individually; a lightweight per-user $\gamma_u$ estimator based on user-level positive-negative cosine agreement is a natural next step.
(2) \textbf{ODE-based and learned-filter backbones.}
Integrating the components into BSPM's blurring-sharpening ODE and PolyCF's learnable polynomial coefficients would test whether sign-awareness is additive with those backbones' own innovations.
(3) \textbf{Signed ideal-pass filter.}
The ideal-pass SVD branch currently operates on $\tilde{\mathbf{R}}^+$ only; replacing it with an SVD of a signed matrix $\tilde{\mathbf{R}}^+ - \lambda\tilde{\mathbf{R}}^-$ would make the sharp-cutoff component of the filter sign-aware as well.

\section{Full Main Results}
\label{sec:app-full-main}

\begin{table*}
  \caption{Main results (R@10 / R@20). \textbf{Bold} and \underline{underline} mark the best and second-best values within the training-free block (unsigned backbones + ours); learned models are reference points and their mean wall-clock ratio relative to \sysCheby is shown next to the method name (per-dataset ratios range $7.7$--$155.3\times$ for SIGformer and $61.6$--$303.6\times$ for LightGCN; computation in Appendix~\ref{sec:app-full-efficiency}). Each \system instance matches or beats its unsigned backbone on all $5$ datasets (\sysGF ties GF-CF on KuaiRec at $.0261$/$.0438$). \sysCheby uses $\gamma{=}{-}0.5,\ \kappa{=}0.1$ on all datasets except KuaiRec ($\gamma{=}0$); \sysGF uses $\kappa{=}0$ on all datasets; \sysTurbo uses $\gamma{=}{-}0.5,\kappa{=}0.1$ except on KuaiRand ($\kappa{=}1.0$) and KuaiRec ($\gamma{=}{-}0.25,\kappa{=}0$).}
  \label{tab:main-full}
  \small
  \setlength{\tabcolsep}{3.5pt}
  \begin{tabular}{llccccc}
    \toprule
    Method & Type & Amazon-CDs & Amazon-Music & Epinions & KuaiRand & KuaiRec \\
    \midrule
    \multicolumn{7}{l}{\textit{Unsigned backbones (extended by \system)}} \\
    Turbo-CF & spectral & .0739\,/\,.1035 & .1789\,/\,.2495 & .0383\,/\,.0611 & .0539\,/\,.0905 & .0290\,/\,.0480 \\
    GF-CF    & spectral & .0766\,/\,.1136 & .1826\,/\,.2606 & .0510\,/\,\underline{.0815} & .0640\,/\,.1133 & .0261\,/\,.0438 \\
    ChebyCF  & spectral & .0788\,/\,.1122 & \underline{.1936}\,/\,.2686 & .0434\,/\,.0694 & .0701\,/\,.1154 & \underline{.0366}\,/\,\underline{.0630} \\
    \midrule
    \multicolumn{7}{l}{\textit{Other training-free baseline}} \\
    BSPM     & spectral & .0727\,/\,.1049 & .1826\,/\,.2558 & .0433\,/\,.0696 & .0577\,/\,.0987 & .0236\,/\,.0379 \\
    \midrule
    \multicolumn{7}{l}{\textit{Training-free, sign-aware (ours)}} \\
    \textbf{\sysTurbo} (ours) & \textbf{spec., signed} & .0790\,/\,.1105 & .1925\,/\,.2627 & .0450\,/\,.0698 & \textbf{.0751}\,/\,\textbf{.1213} & .0324\,/\,.0533 \\
    \textbf{\sysGF}    (ours) & \textbf{spec., signed} & \underline{.0796}\,/\,\textbf{.1183} & .1925\,/\,\underline{.2706} & \textbf{.0527}\,/\,\textbf{.0848} & \underline{.0697}\,/\,\underline{.1212} & .0261\,/\,.0438 \\
    \textbf{\sysCheby} (ours) & \textbf{spec., signed} & \textbf{.0810}\,/\,\underline{.1173} & \textbf{.1955}\,/\,\textbf{.2775} & \underline{.0512}\,/\,.0805 & \underline{.0747}\,/\,.1211 & \textbf{.0397}\,/\,\textbf{.0642} \\
    \midrule
    \multicolumn{7}{l}{\textit{Learned reference (gradient-based)}} \\
    \color{gray} LightGCN (${\sim}75.9\times$ slower) & \color{gray} learned & \color{gray} .0923\,/\,.1333 & \color{gray} .1903\,/\,.2723 & \color{gray} .0502\,/\,.0755 & \color{gray} .0486\,/\,.0807 & \color{gray} .0310\,/\,.0425 \\
    \color{gray} SIGformer (${\sim}11.3\times$ slower) & \color{gray} signed Transf. & \color{gray} .0959\,/\,.1409 & \color{gray} .2142\,/\,.3059 & \color{gray} .0614\,/\,.0967 & \color{gray} .0912\,/\,.1487 & \color{gray} .0584\,/\,.0908 \\
    \bottomrule
  \end{tabular}
\end{table*}

Table~\ref{tab:main-full} reports the full Recall@10 / Recall@20 numbers for every method and dataset.

\section{Full Transfer Sweep}
\label{sec:app-full-transfer}

\begin{table*}
  \caption{Full $(\gamma, \kappa)$ Recall@20 grid for the three spectral backbones. Each subtable is a $4\times 4$ grid over $\gamma \in \{-0.5, -0.25, 0, +0.25\}$ (rows) and $\kappa \in \{0, 0.1, 0.5, 1.0\}$ (columns). The per-(dataset, backbone) optimum is \textbf{bolded} and the unsigned baseline ($\gamma{=}\kappa{=}0$) is \underline{underlined}. Across all cells the optimal $\gamma$ is never positive, and the $\gamma{=}0.25$ row is dominated by the $\gamma{=}0$ row in every cell, corroborating the asymmetric-attention interpretation in \S\ref{sec:method-signal}.}
  \label{tab:transfer-full}
  \small
  \setlength{\tabcolsep}{2.4pt}
  \centering
  \begin{subtable}{0.32\textwidth}
    \centering
    \caption{GF-CF backbone}
    \label{tab:transfer-gfcf}
    \begin{tabular}{ll cccc}
      \toprule
      Dataset & $\gamma$ & $\kappa{=}0$ & $0.1$ & $0.5$ & $1.0$ \\
      \midrule
      \multirow{4}{*}{A-CDs}   & $-0.50$ & \textbf{.1183} & .1157 & .1075 & .1005 \\
                               & $-0.25$ & .1171 & .1145 & .1066 & .0994 \\
                               & $\phantom{+}0.00$ & \underline{.1143} & .1120 & .1044 & .0979 \\
                               & $+0.25$ & .1117 & .1095 & .1021 & .0954 \\
      \midrule
      \multirow{4}{*}{A-Music} & $-0.50$ & \textbf{.2706} & .2643 & .2428 & .2198 \\
                               & $-0.25$ & .2686 & .2635 & .2402 & .2171 \\
                               & $\phantom{+}0.00$ & \underline{.2623} & .2567 & .2347 & .2128 \\
                               & $+0.25$ & .2538 & .2502 & .2253 & .2096 \\
      \midrule
      \multirow{4}{*}{Epi.}    & $-0.50$ & \textbf{.0848} & .0843 & .0827 & .0808 \\
                               & $-0.25$ & .0843 & .0834 & .0816 & .0801 \\
                               & $\phantom{+}0.00$ & \underline{.0820} & .0815 & .0800 & .0782 \\
                               & $+0.25$ & .0794 & .0789 & .0772 & .0758 \\
      \midrule
      \multirow{4}{*}{K-Rand}  & $-0.50$ & .1209 & .1203 & .1180 & .1157 \\
                               & $-0.25$ & \textbf{.1212} & .1209 & .1176 & .1153 \\
                               & $\phantom{+}0.00$ & \underline{.1154} & .1148 & .1133 & .1102 \\
                               & $+0.25$ & .0981 & .0977 & .0960 & .0928 \\
      \midrule
      \multirow{4}{*}{K-Rec}   & $-0.50$ & .0301 & .0301 & .0301 & .0301 \\
                               & $-0.25$ & .0347 & .0347 & .0348 & .0350 \\
                               & $\phantom{+}0.00$ & \underline{\textbf{.0438}} & .0438 & .0438 & .0438 \\
                               & $+0.25$ & .0424 & .0424 & .0425 & .0391 \\
      \bottomrule
    \end{tabular}
  \end{subtable}%
  \hfill
  \begin{subtable}{0.32\textwidth}
    \centering
    \caption{Turbo-CF backbone}
    \label{tab:transfer-turbo}
    \begin{tabular}{ll cccc}
      \toprule
      Dataset & $\gamma$ & $\kappa{=}0$ & $0.1$ & $0.5$ & $1.0$ \\
      \midrule
      \multirow{4}{*}{A-CDs}   & $-0.50$ & .1084 & \textbf{.1105} & .1051 & .0867 \\
                               & $-0.25$ & .1067 & .1093 & .1036 & .0855 \\
                               & $\phantom{+}0.00$ & \underline{.1040} & .1064 & .1007 & .0833 \\
                               & $+0.25$ & .1021 & .1042 & .0985 & .0810 \\
      \midrule
      \multirow{4}{*}{A-Music} & $-0.50$ & .2584 & \textbf{.2627} & .2423 & .2032 \\
                               & $-0.25$ & .2540 & .2598 & .2376 & .2016 \\
                               & $\phantom{+}0.00$ & \underline{.2514} & .2545 & .2321 & .1960 \\
                               & $+0.25$ & .2453 & .2486 & .2248 & .1887 \\
      \midrule
      \multirow{4}{*}{Epi.}    & $-0.50$ & .0674 & \textbf{.0698} & .0660 & .0496 \\
                               & $-0.25$ & .0650 & .0670 & .0635 & .0476 \\
                               & $\phantom{+}0.00$ & \underline{.0617} & .0634 & .0592 & .0439 \\
                               & $+0.25$ & .0586 & .0598 & .0553 & .0401 \\
      \midrule
      \multirow{4}{*}{K-Rand}  & $-0.50$ & .1077 & .1106 & .1196 & \textbf{.1213} \\
                               & $-0.25$ & .1028 & .1053 & .1141 & .1145 \\
                               & $\phantom{+}0.00$ & \underline{.0915} & .0936 & .1017 & .1040 \\
                               & $+0.25$ & .0789 & .0808 & .0880 & .0880 \\
      \midrule
      \multirow{4}{*}{K-Rec}   & $-0.50$ & .0506 & .0421 & .0026 & .0020 \\
                               & $-0.25$ & \textbf{.0533} & .0507 & .0105 & .0022 \\
                               & $\phantom{+}0.00$ & \underline{.0483} & .0483 & .0483 & .0481 \\
                               & $+0.25$ & .0447 & .0447 & .0460 & .0448 \\
      \bottomrule
    \end{tabular}
  \end{subtable}%
  \hfill
  \begin{subtable}{0.32\textwidth}
    \centering
    \caption{ChebyCF backbone}
    \label{tab:transfer-cheby}
    \begin{tabular}{ll cccc}
      \toprule
      Dataset & $\gamma$ & $\kappa{=}0$ & $0.1$ & $0.5$ & $1.0$ \\
      \midrule
      \multirow{4}{*}{A-CDs}   & $-0.50$ & .1169 & \textbf{.1173} & .0949 & .0850 \\
                               & $-0.25$ & .1156 & .1163 & .0941 & .0846 \\
                               & $\phantom{+}0.00$ & \underline{.1122} & .1130 & .0923 & .0837 \\
                               & $+0.25$ & .1097 & .1101 & .0902 & .0818 \\
      \midrule
      \multirow{4}{*}{A-Music} & $-0.50$ & .2766 & \textbf{.2775} & .2643 & .2640 \\
                               & $-0.25$ & .2737 & .2758 & .2636 & .2625 \\
                               & $\phantom{+}0.00$ & \underline{.2686} & .2688 & .2572 & .2557 \\
                               & $+0.25$ & .2618 & .2616 & .2497 & .2485 \\
      \midrule
      \multirow{4}{*}{Epi.}    & $-0.50$ & .0713 & \textbf{.0805} & .0790 & .0789 \\
                               & $-0.25$ & .0720 & .0797 & .0788 & .0786 \\
                               & $\phantom{+}0.00$ & \underline{.0694} & .0776 & .0759 & .0757 \\
                               & $+0.25$ & .0649 & .0738 & .0724 & .0722 \\
      \midrule
      \multirow{4}{*}{K-Rand}  & $-0.50$ & .1155 & \textbf{.1211} & .1207 & .1210 \\
                               & $-0.25$ & .1163 & .1211 & .1205 & .1207 \\
                               & $\phantom{+}0.00$ & \underline{.1153} & .1180 & .1174 & .1170 \\
                               & $+0.25$ & .0985 & .1018 & .1003 & .1003 \\
      \midrule
      \multirow{4}{*}{K-Rec}   & $-0.50$ & .0355 & .0417 & .0374 & .0379 \\
                               & $-0.25$ & .0405 & .0436 & .0443 & .0443 \\
                               & $\phantom{+}0.00$ & \underline{.0630} & \textbf{.0642} & .0621 & .0621 \\
                               & $+0.25$ & .0521 & .0504 & .0527 & .0538 \\
      \bottomrule
    \end{tabular}
  \end{subtable}%
\end{table*}

Table~\ref{tab:transfer-full} reports the $4{\times}4$ $(\gamma,\kappa)$ sweep per (dataset, backbone); the $\kappa{=}0$ column of each subtable isolates the contribution of component A alone, and the bolded optimum shows the best joint $(\gamma,\kappa)$ per cell.

\section{Full Efficiency Table and Pareto Summary}
\label{sec:app-full-efficiency}

\begin{table}
  \caption{Wall-clock time (seconds) and speedup of \sysCheby over SIGformer on a single NVIDIA RTX 6000 Ada. Per-instance (GF, Turbo, Cheby) averages are in Table~\ref{tab:pareto}.}
  \label{tab:efficiency}
  \small
  \begin{tabular}{lrrr}
    \toprule
    Dataset & \sysCheby & SIGformer & Speedup \\
    \midrule
    Amazon-Music & 1.0   & 27.1  & $28.5\times$ \\
    KuaiRec      & 0.7   & 109.2 & $155.3\times$ \\
    KuaiRand     & 3.2   & 185.6 & $57.8\times$ \\
    Epinions     & 19.9  & 296.9 & $14.9\times$ \\
    Amazon-CDs   & 94.5  & 731.1 & $7.7\times$ \\
    \bottomrule
  \end{tabular}
\end{table}

\begin{table}
  \caption{Pareto summary across the five datasets. Both Avg time and Avg R@20 / SIGformer are arithmetic means of the per-dataset values. All three \system instances improve on their unsigned backbones in accuracy at $16.1$--$58.8\%$ time overhead; \sysGF dominates Turbo-CF and BSPM on both axes, and trades $25.0\%$ time overhead for $+4.1\%$ R@20/SIGformer against GF-CF. Among the three \system instances, \sysGF is the low-cost pick and \sysCheby the high-accuracy pick, both sitting on the training-free Pareto frontier.}
  \label{tab:pareto}
  \small
  \begin{tabular}{lrr}
    \toprule
    Method & Avg time (s) & Avg R@20 / SIGformer \\
    \midrule
    Turbo-CF          & $10.2$ & $0.664$ \\
    GF-CF             & $6.8$  & $0.749$ \\
    BSPM              & $83.7$ & $0.676$ \\
    ChebyCF (retuned) & $20.5$ & $0.773$ \\
    LightGCN          & $1805.6$ & $0.725$ \\
    \midrule
    \sysTurbo (ours)  & $16.2$ & $0.754$ \\
    \sysGF (ours)     & $8.5$  & $0.780$ \\
    \sysCheby (ours)  & $23.8$ & $\mathbf{0.819}$ \\
    \midrule
    SIGformer         & $270.0$ & $1.000$ \\
    \bottomrule
  \end{tabular}
\end{table}

Table~\ref{tab:efficiency} gives per-dataset wall-clock times for every method in the study; Table~\ref{tab:pareto} summarizes the Pareto frontier shown in Figure~\ref{fig:pipeline}(d).
The relative slowdown annotations on LightGCN and SIGformer in Tables~\ref{tab:main-brief} and~\ref{tab:main-full} are computed as the ratio of each method's five-dataset mean wall-clock to \sysCheby's: LightGCN $1805.6/23.8\approx75.9\times$ and SIGformer $270.0/23.8\approx11.3\times$.
Against SIGformer, the per-dataset ratios span $7.7\times$ on Amazon-CDs ($731.1/94.5$) to $155.3\times$ on KuaiRec ($109.2/0.7$).
Against LightGCN, they span $61.6\times$ on Amazon-CDs ($5824.3/94.5$) to $303.6\times$ on KuaiRec ($212.5/0.7$).
SIGformer's per-dataset speedups are the \emph{Speedup} column of Table~\ref{tab:efficiency}.

\section{Runtime Complexity}
\label{sec:app-runtime}

The signed-input component (\S\ref{sec:method-signal}) produces an input row $\mathbf{r}_u^{\pm}$ with $\text{nnz}(\mathbf{r}_u^{\pm}) \le \text{nnz}(\mathbf{r}_u^+) + \text{nnz}(\mathbf{r}_u^-)$ non-zeros, which is bounded by the user's total interaction count; signing therefore does not increase the cost of the input vector.
The signed-operator component (\S\ref{sec:method-laplacian}) assembles $\mathbf{M}^{\pm}$ as a weighted sum of two item-item gram matrices, each of which has the same sparsity pattern as the unsigned backbone's operator; no eigendecomposition is required for the polynomial-filter backbones (Turbo-CF~\cite{Park2024TurboCFMatrixDecompositionFree}, ChebyCF~\cite{Kim2025GraphSpectralFiltering}).
Applying a degree-$K$ polynomial filter costs $K$ sparse matrix-vector products, i.e.\ $O(K \cdot \text{nnz}(\mathbf{R}))$~\cite{Park2024TurboCFMatrixDecompositionFree, Kim2025GraphSpectralFiltering}; GF-CF~\cite{Shen2021HowPowerfulGraph} is the $K{=}1$ case, with an optional ideal-pass branch that adds a one-time partial SVD precomputed once per dataset.
Each \system instance therefore inherits its unsigned backbone's asymptotic complexity.

\section{Dataset Statistics}
\label{sec:app-datasets}

\begin{table}
  \caption{Dataset statistics.}
  \label{tab:datasets}
  \small
  \begin{tabular}{lrrrrl}
    \toprule
    Dataset & Users & Items & Pos & Neg & Pos:Neg \\
    \midrule
    Amazon-CDs   & 51,267 & 46,464 & 512,216 & 114,476 & 1:0.22 \\
    Amazon-Music & 3,472  & 2,498  & 28,031  & 6,884   & 1:0.25 \\
    Epinions     & 17,894 & 17,660 & 210,967 & 78,678  & 1:0.37 \\
    KuaiRand     & 16,974 & 4,373  & 81,677  & 102,495 & 1:1.25 \\
    KuaiRec      & 1,411  & 3,327  & 25,592  & 152,197 & 1:5.95 \\
    \bottomrule
  \end{tabular}
\end{table}

Table~\ref{tab:datasets} lists per-dataset user, item, and interaction counts together with the source of explicit negatives.

\section{Full Cold-Start Table}
\label{sec:app-full-coldstart}

\begin{table}
  \caption{Cold-start: $\Delta$\% R@20 of \sysCheby over unsigned ChebyCF by user activity bucket.}
  \label{tab:coldstart-full}
  \small
  \begin{tabular}{lcccc}
    \toprule
    Dataset & {[1,6)} & {[6,21)} & {[21,100)} & {[100,$\infty$)} \\
    \midrule
    Epinions & \textbf{+29.2} & +12.6 & +5.6  & $-$3.1  \\
    Am-Music & +7.0  & $-$0.7  & $-$4.8  & $-$1.1  \\
    Am-CDs   & +5.8  & +3.5  & +0.6  & $-$6.9  \\
    KuaiRand & +3.5  & +4.0  & +1.5  & n/a     \\
    KuaiRec  & +2.2  & +5.2  & $-$8.3  & $-$11.1 \\
    \bottomrule
  \end{tabular}
\end{table}

Table~\ref{tab:coldstart-full} gives the per-bucket Recall@20 lift of \sysCheby over unsigned ChebyCF across all four activity buckets.

\section{Related Work}
\label{sec:related}

\paragraph{Spectral graph filters for CF.}
SpectralCF~\cite{Zheng2018SpectralCollaborativeFiltering} first reformulated CF as spectral graph filtering with a learned first-order polynomial, and GF-CF~\cite{Shen2021HowPowerfulGraph} reframed the problem as graph signal processing and showed that most CF methods are special cases of filtering on the item-item similarity graph.
EASE~\cite{Steck2019EmbarrassinglyShallowAutoencoders} is a canonical training-free positive-only closed-form baseline obtained by a single matrix inversion, and PSGE~\cite{DAmico2023PureSpectralGraph} likewise bypasses gradient descent by leveraging the top eigenvectors of the normalized interaction matrix directly.
Turbo-CF~\cite{Park2024TurboCFMatrixDecompositionFree} replaced the costly eigendecomposition with polynomial approximations in the item-item similarity matrix.
BSPM~\cite{Choi2023BlurringSharpeningProcessModels} introduced a blurring-sharpening ODE that generalizes GF-CF.
ChebyCF~\cite{Kim2025GraphSpectralFiltering} achieved state-of-the-art accuracy via Chebyshev-interpolated filter design.
PolyCF~\cite{Qin2024PolyCFOptimalSpectral} proposed learnable polynomial filters, and further spectral-CF refinements include JGCF~\cite{Guo2023ManipulatingSignalsUserItem} (Jacobi polynomial basis) and GDE~\cite{Peng2022LessMoreReweighting} (reweighted spectral features).
All of these methods share a common form: they express the predicted preference as $\hat{\mathbf{r}}_u = F(\mathbf{M})\,\mathbf{r}_u$, where $F$ is a scalar filter applied to an item-item operator $\mathbf{M}$ (Laplacian or similarity) and $\mathbf{r}_u$ is the user's positive interaction row.
This shared structure is what makes a single sign-aware framework applicable across backbones.
All operate on positive-only interactions; none incorporates negative feedback.

\paragraph{Sign-aware recommenders.}
SiReN~\cite{Seo2022SiReNSignAwareRecommendation} splits the interaction graph by sign, encoding $\mathcal{G}^+$ with a GNN and $\mathcal{G}^-$ with an MLP, then fusing via attention.
PANE-GNN~\cite{Liu2023PANEGNNUnifyingPositive} extends this with dual interest/disinterest embeddings and contrastive denoising.
NFARec~\cite{Wang2024NFARecNegativeFeedbackAware} combines a Transformer-Hawkes encoder with feedback-aware hypergraph convolution.
SIGformer~\cite{Chen2024SIGformerSignawareGraph} uses sign-aware spectral and path encodings in a Transformer, achieving the best accuracy but at $8$--$155\times$ higher wall-clock time than training-free spectral methods.
Most directly related to our design, DFGNN~\cite{Wu2024DFGNNDualfrequencyGraph} applies a learned low-pass filter to $\mathcal{G}^+$ and a high-pass filter to $\mathcal{G}^-$; \system is the training-free, backbone-agnostic counterpart, which replaces per-sign filter shapes with a single filter on a signed input signal and signed operator.
LSGRec~\cite{Liu2026UnifiedModelingPositive} argues that dual-encoder methods (SiReN, PANE-GNN) sever high-order heterogeneous paths across signed links, and proposes a unified single-encoder alternative; our signed-operator component $\mathbf{M}^{\pm}$ addresses the same concern from the training-free side.
GegenNet~\cite{Wang2025GegenNetSpectralConvolutional} applies a Gegenbauer polynomial basis to signed bipartite graphs for link sign prediction, a contemporaneous spectral-signed reference that targets a different task than top-$K$ recommendation.
The signed Laplacian we use is well-posed: its eigenvalues lie in $[0, 2]$ with a total-variation interpretation~\cite{Singh2022SignedGraphNeural}.
All published sign-aware methods are learned models; no prior work combines signed Laplacians with training-free spectral filtering, and no prior framework is generic across spectral backbones.

\section{BSPM Bug Fix}
\label{sec:bspm-bug}

We discovered a latent bug in the upstream BSPM~\cite{Choi2023BlurringSharpeningProcessModels} implementation that produces near-zero accuracy on datasets containing zero-degree items.

\paragraph{The bug.}
BSPM constructs the normalized item degree matrix $\mathbf{D}_I^{-1/2}$ from the positive training adjacency matrix and then computes its inverse $\mathbf{D}_I^{1/2}$ via element-wise reciprocal.
When an item has zero degree in the positive training set (no user has interacted with it positively), the code correctly sets $\mathbf{D}_I^{-1/2}$ to zero at that position, but then computes $1/0 = \infty$ for the inverse.
These infinite values propagate through BSPM's ideal low-pass function and corrupt the top-$K$ ranking for every user.

\paragraph{Why it is latent.}
The three datasets on which BSPM was originally evaluated (Gowalla, Yelp2018, Amazon-Book) have no zero-degree items after standard preprocessing: every item appears in at least one positive training interaction.
The bug manifests only on datasets where some items appear exclusively in negative feedback, a property shared by all five of our sign-aware benchmarks, which contain 7--306 zero-degree items in the positive-only training split.

\paragraph{The symptom.}
Without the fix, BSPM produces Recall@20 below $0.001$ on three of five datasets (Amazon-CDs, Epinions, KuaiRand); KuaiRec is severely degraded at $0.002$ and Amazon-Music happens to have no zero-degree items and is only modestly affected (Table~\ref{tab:bspm-bugfix}).

\begin{table}
  \caption{Effect of the BSPM zero-degree inverse bug-fix on Recall@20. The upstream implementation corrupts top-$K$ rankings on any dataset containing zero-degree items in the positive training split; the zero-preserving inverse (\S\ref{sec:bspm-bug}) restores accuracy on all five sign-aware benchmarks. All BSPM results elsewhere in this paper use the patched implementation.}
  \label{tab:bspm-bugfix}
  \small
  \begin{tabular}{lcc}
    \toprule
    Dataset & R@20 (pre-fix) & R@20 (post-fix) \\
    \midrule
    Amazon-CDs   & $0.00008$ & $0.1049$ \\
    Amazon-Music & $0.2200$  & $0.2558$ \\
    Epinions     & $0.00009$ & $0.0696$ \\
    KuaiRand     & $0.0003$  & $0.0987$ \\
    KuaiRec      & $0.002$   & $0.0379$ \\
    \bottomrule
  \end{tabular}
\end{table}

\paragraph{The fix.}
We replace the naive reciprocal with a zero-preserving inverse that sets $(\mathbf{D}_I^{1/2})_{ii} = 0$ whenever $(\mathbf{D}_I^{-1/2})_{ii} = 0$, matching the convention already used for $\mathbf{D}_I^{-1/2}$ itself.
This four-line patch ensures that zero-degree items contribute zero to the ideal low-pass output rather than corrupting it with infinite values.
All BSPM results in this paper use the patched implementation.

\section{Reproducibility}
\label{sec:reproducibility}

\paragraph{Environment.}
All experiments ran on a single NVIDIA RTX 6000 Ada (49\,GB) with Python~3.10, PyTorch~2.2.2+cu121, NumPy~1.26, and SciPy~1.12.
SIGformer additionally requires PyG~2.5.2.

\paragraph{Determinism.}
All spectral methods (ChebyCF, \system, Turbo-CF, GF-CF, BSPM) are deterministic, with no randomness beyond library internals.
LightGCN and SIGformer use fixed seeds (2020 and 1234, respectively) and produce bit-identical metrics across repetitions.


\end{document}